\documentclass[10pt,journal,compsoc]{IEEEtran}

\usepackage{epsfig,endnotes}
\usepackage{amssymb}
\usepackage{graphicx}
\usepackage{amsfonts}
\usepackage{dsfont}
\usepackage{amsmath}
\usepackage{color}
\usepackage{booktabs}
\usepackage{epstopdf}
\usepackage{url}
\usepackage{tabularx}
\usepackage{verbatim} % This is for comment and to be del later
\usepackage{setspace}
\usepackage{caption}

\newtheorem{definition}{Definition}
\newtheorem{theorem}{Theorem}

\begin{document}

\title{On the Security of MTA-OTIBASs (Multiple-TA
One-Time Identity-Based Aggregate Signatures)}

\author{Lei Zhang,~\IEEEmembership{Member,~IEEE,} Qianhong Wu,~\IEEEmembership{Member,~IEEE,}
        Josep Domingo-Ferrer,~\IEEEmembership{Fellow,~IEEE} \\Bo Qin, Chuanyan Hu
         % <-this % stops a space
\IEEEcompsocitemizethanks{\IEEEcompsocthanksitem Lei Zhang and Chuanyan Hu are with the Shanghai Key Laboratory of Trustworthy Computing, East China Normal University, and the State Key Laboratory of Integrated Services Networks, Xidian University; Qianhong Wu is with the School of Electronic and Information Engineering, Beihang University, and the State Key Laboratory of Information Security (Institute of Information Engineering, Chinese Academy of Sciences); Josep Domingo-Ferrer is with the Department of Computer Engineering and Mathematics, Universitat Rovira i Virgili; Bo Qin is with the Key Laboratory of Data Engineering and Knowledge Engineering, Ministry of Education, School of Information, Renmin University of China, the Key Laboratory of Cryptologic Technology and Information Security, Ministry of Education, Shandong University, and the State Key Laboratory of Cryptology (e-mail: leizhang@sei.ecnu.edu.cn,
qhwu@xidian.edu.cn, josep.domingo@urv.cat, bo.qin@ruc.edu.cn, chuanyanhu@ecnu.edu.cn).
%\IEEEcompsocthanksitem
}% <-this % stops a space
\thanks{}}

%\markboth{Journal of \LaTeX\ Class Files,~Vol.~6, No.~1, January~2007}%
%{Shell \MakeLowercase{\textit{et al.}}: Bare Demo of IEEEtran.cls for Computer Society Journals}

\IEEEcompsoctitleabstractindextext{%
\begin{abstract}
In~\cite{IEEE-T-ITS} the authors proposed a new
aggregate signature scheme
referred to as multiple-TA (trusted authority) one-time identity-based aggregate signature (MTA-OTIBAS). Further, they gave a concrete MTA-OTIBAS scheme.
We recall here the definition of MTA-OTIBAS and the concrete proposed scheme.
Then we prove that
our MTA-OTIBAS concrete scheme is existentially
unforgeable against adaptively chosen-message attacks in
the random oracle model under the co-CDH problem
assumption.
\end{abstract}

}

\maketitle

\IEEEdisplaynontitleabstractindextext

\IEEEpeerreviewmaketitle

\section{Introduction}

%Identity-based cryptosystems~\cite{Shamir} have been suggested to eliminate the certificate management problem. In such
%cryptosystems, the public key of an entity is just its identity and
%the private key of the entity is generated by the trusted authority
%(TA) using its master secret key. An aggregate signature scheme \cite{Gentry,ZWQD11} allows an efficient algorithm to aggregate signatures on distinct
%messages from different users into one single signature. The resulting aggregate signature can convince a verifier that the signature was indeed produced by those users. This paper studies aggregate signatures in identity-based
%cryptosystems with multiple TAs.
%
%In this paper, we first propose a new notion referred to as MTA-OTIBAS. An MTA-OTIBAS scheme consists of a root TA,
%several lower-level TAs and users. Each lower-level TA is enrolled
%by the root TA. A user can register to any lower-level TA and
%compute a signature on a message if the user has obtained a private
%key from the lower-level TA. The signature is only valid under the
%user's identity and the public information of the lower-level TA.
In~\cite{IEEE-T-ITS} we proposed a new
aggregate signature scheme
referred to as multiple-TA (trusted authority) one-time
identity-based aggregate signature (MTA-OTIBAS). Further,
we gave a concrete MTA-OTIBAS scheme.
We first recall the notion of MTA-OTIBAS;
we then recall its formal definition and the concrete scheme
proposed in~\cite{IEEE-T-ITS}.
Then, we give the detailed security proof
of MTA-OTIBAS (not given in~\cite{IEEE-T-ITS}).

An MTA-OTIBAS scheme has the following
features. Firstly, each user's public key is his identity, so no
certificate is needed on the public key, which avoids the
certificate management overhead. Secondly, a signer's private key
(corresponding to an identity and a lower-level TA) is restricted to
be used only once; after that, the signer's private key should be
updated. Thirdly,
the MTA-OTIBAS scheme also allows signature
aggregation and fast verification, i.e., $n$ signatures can be
aggregated into a single short signature (even signatures generated
by signers enrolled by different lower-level TAs), which greatly saves storage space, and can be
verified simultaneously.

We recall the formal definition of MTA-OTIBAS in Section~\ref{IBC}.
In Section~\ref{specific} we recall the concrete MTA-OTIBAS scheme.
Then in Section~\ref{correctness and
security} we prove that
our MTA-OTIBAS concrete scheme is existentially
unforgeable against adaptively chosen-message attacks in
the random oracle model under the co-CDH problem
assumption.

%\section{MTA-OTIBAS}\label{OTIBAS}

\section{Definition of MTA-OTIBAS}\label{IBC}

An MTA-OTIBAS scheme consists of six
algorithms, i.e., {\sf Root.Setup}, {\sf LowLevel.Setup}, {\sf
Extract}, {\sf Sign}, {\sf Aggregate}, and {\sf Verify}. {\sf
Root.Setup} is run by the root TA to generate the global system
parameters and system master key. {\sf LowLevel.Setup} is an
interactive protocol run between a lower-level TA and the root TA.
It generates the secret key, public key and certificate of the
lower-level TA. {\sf Extract} takes as input a lower-level TA's
secret key and a signer's identity, and outputs a private key for
the signer. {\sf Sign} takes as input a signer's identity, his private
key, the certificate of the signer's corresponding lower-level TA
and any message, and outputs a signature on the message. The
signature is only valid under the signer's identity and the
certificate of his corresponding lower-level TA. A restriction
here is that a private key corresponding to a specific identity
issued by a lower-level TA can be used only once. However, the same
identity can be enrolled by different lower-level TAs. This implies
that the corruption of a lower-level TA does not influence the signers
enrolled by other lower-level TAs. {\sf Aggregate} is used to
aggregate $n$ message-signature pairs generated by the {\sf Sign}
procedure into a single signature, i.e., an aggregate signature. {\sf
Verify} is used to check the validity of an aggregate signature. It
takes as input $n$ messages, the corresponding aggregate signature,
$n$ identities enrolled by $l$ lower-level TAs, and outputs 1 or 0
to represent whether the aggregate signature is valid or not.

\section{A concrete MTA-OTIBAS scheme}
\label{specific}

Our MTA-OTIBAS scheme is
realized using bilinear
maps which are
widely employed in identity-based cryptosystems. A map
$\hat{e}:\mathbb{G}_1\times \mathbb{G}_2\rightarrow \mathbb{G}_T$ is
called a bilinear map if $\hat{e}(g_1,g_2)\neq 1$ and
$\hat{e}(g_1^\alpha,g_2^\beta)=\hat{e}(g_1,g_2)^{\alpha\beta}$ for
all $\alpha, \beta\in \mathbb{Z}_q^*$, where $\mathbb{G}_1,\mathbb{G}_2$ are
two cyclic groups of prime order $q$, $\mathbb{G}_T$ is a
multiplicative cyclic group of the same order, $g_1$ is a generator
of $\mathbb{G}_1$, and $g_2$ is a generator of $\mathbb{G}_2$. By
exploiting bilinear maps, we implement our MTA-OTIBAS scheme.

\smallskip
{\sf Root.Setup}:
\smallskip
The root TA runs this algorithm to generate the system parameters as
follows:
\begin{enumerate}
  \item Choose $q,\mathbb{G}_1,\mathbb{G}_2,\mathbb{G}_T,g_1,g_2,\hat{e},\psi$,
  where $\psi$ is a computable isomorphism from
$\mathbb{G}_2$ to $\mathbb{G}_1$, with $\psi(g_2) = g_1$ \cite{Zhang10}.

  \item Pick $\kappa\in \mathbb{Z}_q^*$ as its master secret key, and
compute $y=g_2^\kappa$ as its master public key.

  \item Select cryptographic hash functions $H_0(\cdot):\{0,
  1\}^*$ $\rightarrow$ $\mathbb{G}_1$ and $H_1(\cdot):\{0, 1\}^*\rightarrow
  \mathbb{Z}_q^*$.

  \item Publish the system global parameter
$\Psi=(\hat{e}, q, \mathbb{G}_1, \mathbb{G}_2, \mathbb{G}_T, g_1,
g_2, H_0(\cdot), H_1(\cdot),\psi)$.
\end{enumerate}

\smallskip
{\sf LowerLevel.Setup}: In an MTA-OTIBAS scheme, before a lower-level TA can recruit members, it must be
enrolled by the root TA. The root TA may add the public information
of a lower-level TA (e.g., identity and public key) to the system
global parameters. Let the identity of a lower-level TA
$\mathcal{T}_i$ be $ID_{\mathcal{T}_i}$. $\mathcal{T}_i$ picks
$\kappa_i\in \mathbb{Z}_q^*$ as its secret key and computes
$y_i=g_2^{\kappa_i}$ as its public key. $(ID_{\mathcal{T}_i},y_i)$
are submitted to the root TA. On input $(ID_{\mathcal{T}_i},y_i)$,
the root TA generates a certificate $cert_{\mathcal{T}_i}$ which is
signed using its master secret key. Finally, $cert_{\mathcal{T}_i}$
is sent to $\mathcal{T}_i$.

\smallskip
{\sf Extract}: Suppose a signer with identity $ID_j$ wants to join
the system maintained by $\mathcal{T}_i$ whose secret key is
$\kappa_i$. On input the signer's identity $ID_j$, $\mathcal{T}_i$
generates the private key for the signer as follows:
\begin{enumerate}
  \item Compute $id_{j,0}=H_0(ID_j,0),id_{j,1}=H_0(ID_j,1)$;
  \item Compute $s_{j,i,0} =id_{j,0}^{\kappa_i},s_{j,i,1} =id_{j,1}^{\kappa_i}$, and set $s_{j,i}=(s_{j,i,0},s_{j,i,1})$ as the private key of the signer.
\end{enumerate}

\smallskip
{\sf Sign}:
\smallskip
To sign a message $m_k$, a signer with identity $ID_j$ enrolled by
$\mathcal{T}_i$ and private key $s_{j,i}=(s_{j,i,0},s_{j,i,1})$
computes
$h_k=H_1(m_k,ID_j,cert_{\mathcal{T}_i}),\sigma_k=s_{j,i,0}s_{j,i,1}^{h_k}.$
The signer outputs $\sigma_k$ as the signature on $m_k$.

\smallskip
{\sf Aggregate}:
\smallskip
This publicly computable algorithm aggregates $n$ signatures into a
single signature. Let an entity collect $n$ message-signature pairs
$\{(m_1, \sigma_1), \cdots, (m_n, \sigma_n)\}$ signed by $n$ users
with corresponding identities $\{ID_1, \cdots, ID_n\}$ enrolled by
$l$ lower-level TAs $\{\mathcal{T}_1, \cdots, \mathcal{T}_l\}$. For
simplicity, we assume $\{ID_1,...,ID_{t_1}\}$,
$\{ID_{t_1+1},...,ID_{t_2}\} ,...,\{ID_{t_{l-1}+1},...,ID_{t_l}\}$
are enrolled by $\mathcal{T}_1, \cdots, \mathcal{T}_l$ respectively.
The message-signature pairs are divided into $l$ sets corresponding
to the $l$ lower-level TAs. This algorithm outputs $\Omega$ as the
resulting aggregate signature, where $\Omega=\prod_{i=1}^n
\sigma_i$.

\smallskip
{\sf Verify}:
\smallskip
To verify an aggregate signature $\Omega$ on messages $\{m_1,...,$
$m_n\}$ under $\mathbb{I}_1=\{ID_1,...,ID_{t_1}\}$,
$\mathbb{I}_2=\{ID_{t_1+1},...,ID_{t_2}\}
,...,\mathbb{I}_l=\{ID_{t_{l-1}+1},...,ID_n\}$ enrolled by
$\mathcal{T}_1, \cdots, \mathcal{T}_l$ respectively, the verifier
performs the following steps:
    \begin{enumerate}
      \item For $1\leq j\leq n$, compute $h_j=H_1(m_j,ID_j,cert_{\mathcal{T}_i})$ and
      $id_{j,0}=H_0(ID_j,0)$, $id_{j,1}$ $=H_0(ID_j,1)$.

      \item Define $\mathbb{I}_1'=\{1,...,t_1\}$,
$\mathbb{I}_2'=\{t_1+1,...,t_2\}
,...,\mathbb{I}_l'=\{t_{l-1}+1,...,n\}$. Check
$\hat{e}(\Omega,g_2)\stackrel{?}{=}\prod_{i=1}^l \hat{e}(\prod_{j\in \mathbb{I}_i'}id_{j,0}
      id_{j,1}^{h_j},y_i).$ Output 1 if the equation holds; else output 0.
      \end{enumerate}

\section{Security proof}\label{correctness and
security}

An MTA-OTIBAS scheme should be secure.
Informally, an MTA-OTIBAS scheme is said to be secure if no
polynomial-time attacker not requesting a private key of an entity
enrolled by a lower-level TA can forge an aggregate signature that
is valid
 (i.e., such that {\sf Verify} outputs 1) corresponding to that entity
enrolled by the lower-level TA.

In general, the security of an MTA-OTIBAS scheme is modeled via the following EUF-CMA (existential
universal forgery under adaptive chosen-message attack) game
\cite{Gentry} and takes place between a challenger $\mathcal {CH}$
and an adversary $\mathcal {A}$. The game has the following three
stages:

\smallskip
\noindent {\em Initialize}: $\mathcal {CH}$ runs the {\sf Root.Setup}
algorithm to obtain a master secret key and the system parameters.
$\mathcal {CH}$ then sends the system parameters to $\mathcal {A}$
while keeping secret the master secret key.

\medskip
\noindent \emph{Attack}: $\mathcal {A}$ can perform a polynomially
bounded number of the following types of queries in an adaptive
manner.
\begin{itemize}
  \item {\sf LowerLevel.Setup} \emph{queries}: $\mathcal {A}$ may
  ask $\mathcal {CH}$ to set up a lower-level TA. On input an
  identity $ID_{\mathcal{T}_i}$ of a lower-level TA,
    $\mathcal {CH}$ generates the secret key and certificate of the lower-level TA.

  \item {\sf Corrupt.LowerLevel} \emph{queries}: $\mathcal {A}$ can request the
    secret key of a lower-level TA $\mathcal{T}_i$. On input
    $ID_{\mathcal{T}_i}$,
    $\mathcal {CH}$
    outputs the corresponding secret key of $\mathcal{T}_i$.

  \item {\sf Extract} \emph{queries}: $\mathcal {A}$ can request the
    private key of an entity with identity $ID_j$ issued by a lower-level TA $\mathcal{T}_i$.
    On input $(ID_j,cert_{\mathcal{T}_i})$,
    $\mathcal {CH}$
    outputs the corresponding private key of the entity.

  \item {\sf Sign} \emph{queries}: $\mathcal {A}$ can
    request an entity's signature on a message $m_k$.
    On receiving a query on $(m_k, ID_j,cert_{\mathcal{T}_i})$,
    $\mathcal {CH}$ generates a valid
    signature $\sigma_j$ on $m_k$ under $(ID_j,cert_{\mathcal{T}_i})$,
    and replies with $\sigma_j$.
\end{itemize}

\noindent \emph{Forgery}: $\mathcal {A}$ outputs $l'$ sets of
identities $\mathbb{I}_1^*=\{ID_1^*,$ $...,ID_{t_1}^*\}$,
$\mathbb{I}_2^*=\{ID_{t_1+1}^*,...,ID_{t_2}^*\}
,...,\mathbb{I}_{l'}^*=\{ID_{t_{l'-1}+1}^*,...,$ $ID_n^*\}$ enrolled by
$l'$ lower-level TAs with certificates from the set
$\{cert_{\mathcal{T}_1}^*,...,cert_{\mathcal{T}_{l'}}^*\}$, a set of
$n$ messages $\{m_1^*,...,m_n^*\}$ and an aggregate signature
$\sigma^*$. For simplicity, we assume $m_i^*$ corresponds to
$ID_i^*$ for $i\in \{1,...,n\}$.

\smallskip\noindent $\mathcal {A}$ wins the above game, if all of the
following conditions are satisfied:
\begin{enumerate}
  \item $\sigma^*$ is a valid aggregate signature on messages $\{m_1^*,...,m_n^*\}$
     under $\mathbb{I}_1^*=\{ID_1^*,...,ID_{t_1}^*\}$,
$\mathbb{I}_2^*=\{ID_{t_1+1}^*,...,ID_{t_2}^*\}
,...,\mathbb{I}_{l'}^*=\{ID_{t_{l'-1}+1}^*,...,ID_n^*\}$ and
$\{cert_{\mathcal{T}_1}^*,...,cert_{\mathcal{T}_{l'}}^*\}$.

  \item At least, one
    private key of an entity issued by a lower-level TA is not queried by $\mathcal {A}$ during the {\sf Extract}
  queries and the lower-level TA is not corrupted. Without loss of generality,
  we assume the identity of the entity is
  $ID_1^*$ and its corresponding lower-level TA is
  ${\mathcal{T}_1}^*$ with certificate
  $cert_{\mathcal{T}_1}^*$.

  \item For a message $m\neq m_1^*$, the query $(m, ID_1^*,cert_{\mathcal{T}_1}^*)$
   can be queried at most once,
  and $(m_1^*, ID_1^*,cert_{\mathcal{T}_1}^*)$ is never queried during the {\sf Sign} queries.
\end{enumerate}

We can now define the security of an MTA-OTIBAS scheme in terms of the
above game.

\begin{definition} An MTA-OTIBAS scheme is secure, i.e., secure against existential forgery
under adaptive chosen-message attack, iff the success probability of
any polynomially bounded adversary in the above EUF-CMA game is
negligible.
\end{definition}

We next recall the co-CDH assumption on which the security of the
signature scheme in Section~\ref{specific} rests.

\begin{definition}[co-CDH Assumption]
The co-CDH assumption in two cyclic groups $\mathbb{G}_1$ and
$\mathbb{G}_2$ of prime order $q$ equipped with bilinearity states
that, given $(g_1^a,g_2^b)$ for randomly chosen $a,b\in \mathbb{Z}_q^*$, it
is hard for any polynomial-time algorithm to compute $g_1^{ab}$.
\end{definition}

Regarding the security of our MTA-OTIBAS scheme, we have the following
claim.

\begin{theorem}
\label{teo1} Assume an adversary $\mathcal{A}$ has an advantage
$\epsilon$ in forging an MTA-OTIBAS scheme
of Section~\ref{specific} in an attack modeled by the above
EUF-OTIBAS-CMA game, within a time span $\hat{\tau}$; the adversary
can make at most $q_{H_i}$ times $H_i(\cdot)\ (i=0,1)$ queries,
$q_{L}$ times {\sf LowerLevel.Setup} queries, $q_C$ times {\sf
Corrupt.LowerLevel} queries, $q_E$ times {\sf Extract} queries,
$q_S$ times {\sf Sign} queries. Then the challenger can solve the
co-CDH problem with probability $\epsilon'\geq
\frac{4}{e^2(q_C+q_E+q_S+n+2)^2}\epsilon$ within time
$\hat{\tau}'=\hat{\tau}+\mathcal{O}(4q_{H_0}+q_L+q_S)\tau_{G_1},$
where $\tau_{G_1}$ is the time to compute a point exponentiation in
$\mathbb{G}_1$ and $n$ is the size of the aggregating set.
\end{theorem}

\emph{Proof:}
Let $\mathcal {CH}$ be a co-CDH attacker who receives a co-CDH
challenge instance $(g_1^a,g_2^b)$ and wants to compute the value of
$g_1^{ab}$. $\mathcal {A}$ is an adversary who interacts with
$\mathcal {CH}$ as modeled in the EUF-CMA game. We show how $\mathcal
{C}$ can use $\mathcal {A}$ to break the co-CDH assumption.

\smallskip\noindent $Initialize$: Firstly, $\mathcal {CH}$ selects
$\Psi=(\hat{e}$, $q$, $\mathbb{G}_1$, $\mathbb{G}_2$,
$\mathbb{G}_T$, $g_1$, $g_2$, $y$, $H_0(\cdot)$, $H_1(\cdot),\psi)$,
where $y=g_2^\kappa$, and $\kappa$ is the master secret key; then $\Psi$
is sent to $\mathcal {A}$.

\medskip
\noindent $Attack$:  We consider the hash functions $H_0(\cdot)$ and
$H_1(\cdot)$ as random oracles. $\mathcal {A}$ can perform the
following types of queries in an adaptive manner.

\smallskip \noindent $H_0(\cdot)$ \emph{queries}: $\mathcal {CH}$ maintains a list
$H_0^{list}$ of tuples
$(ID_i,\alpha_{i,0},\alpha_{i,0}',\alpha_{i,1},\alpha_{i,1}',id_{i,0},id_{i,1},coin_i).$
This list is initially empty. Whenever $\mathcal {CH}$ receives an
$H_1$ query on $(ID_i,j)$ (where $j=0$ or 1), $\mathcal {CH}$  does
the following:
\begin{itemize}
  \item If $ID_i$ exists in a previous query,
  find
  $(ID_i,\alpha_{i,0},\alpha_{i,0}',$ $\alpha_{i,1},\alpha_{i,1}',id_{i,0},id_{i,1},coin_i)$
  on $H_1^{list}$ and return $id_{i,j}$.
  \item Else, first flip a coin
$coin_i\in \{0,1\}$ that yields 1 with probability $\delta$ and 0
with probability $1-\delta$. Then do:
      \begin{itemize}
        \item If $coin_i=0$, select $\alpha_{i,0},\alpha_{i,1}\in
        \mathbb{Z}_q^*$, compute
        $id_{i,0}=g_1^{\alpha_{i,0}},id_{i,1}=g_1^{\alpha_{i,1}}$, set
        $\alpha_{i,0}'=\alpha_{i,1}'=0$, return $id_{i,j}$
        and add
        $(ID_i,\alpha_{i,0},$ $\alpha_{i,0}',\alpha_{i,1},\alpha_{i,1}',id_{i,0},id_{i,1},coin_i)$
        to $H_0^{list}$.

        \item Else randomly select $\alpha_{i,0},\alpha_{i,0}',\alpha_{i,1},\alpha_{i,1}'\in
        \mathbb{Z}_q^*$,
         set $id_{i,0}=g_1^{\alpha_{i,0}}{g_1^a}^{\alpha_{i,0}'},id_{i,1}=g_1^{\alpha_{i,1}}{g_1^a}^{\alpha_{i,1}'},$
         and add
         $(ID_i,\alpha_{i,0},\alpha_{i,0}',\alpha_{i,1},\alpha_{i,1}',id_{i,0},id_{i,1},coin_i)$
        to $H_0^{list}$. Return $id_{i,j}$ as the answer.
      \end{itemize}
\end{itemize}

\smallskip
\noindent {\sf LowerLevel.Setup} \emph{queries}: $\mathcal {CH}$
maintains a list $TA^{list}$ of tuples
$(ID_{\mathcal{T}_i},\kappa_i,y_i,cert_{\mathcal{T}_i},coin_{\mathcal{T}_i}).$
On input an identity $ID_{\mathcal{T}_i}$ of a lower-level TA,
$\mathcal {CH}$ does the following:
\begin{itemize}
  \item If there is a tuple $(ID_{\mathcal{T}_i},\kappa_i,y_i,cert_{\mathcal{T}_i},coin_{\mathcal{T}_i})$
  on $TA^{list}$, return $cert_{\mathcal{T}_i}$ as the answer.
  \item Else, choose
$\kappa_i\in \mathbb{Z}_q^*$, flip a coin $coin_{\mathcal{T}_i}\in
\{0,1\}$ that yields 1 with probability $\delta$ and 0 with
probability $1-\delta$ and do the following:
\begin{itemize}
  \item If
$coin_{\mathcal{T}_i}=0$, set $\kappa_i$ as the secret key, compute
$y_i=g_2^{\kappa_i}$, generate a certificate $cert_{\mathcal{T}_i}$
corresponding to $(ID_{\mathcal{T}_i},y_i)$, add
$(ID_{\mathcal{T}_i},\kappa_i,y_i,cert_{\mathcal{T}_i},coin_{\mathcal{T}_i})$
to $TA^{list}$.
  \item Else, compute $y_i=g_2^{b\kappa_i}$, generate a certificate $cert_{\mathcal{T}_i}$
corresponding to $(ID_{\mathcal{T}_i},y_i)$, add
$(ID_{\mathcal{T}_i},\kappa_i,y_i,cert_{\mathcal{T}_i},coin_{\mathcal{T}_i})$
to $TA^{list}$.
\end{itemize}
\end{itemize}

In the rest of this paper, we assume that if a certificate
$coin_{\mathcal{T}_i}$ appears, $\mathcal {A}$ has already made a
corresponding {\sf LowerLevel.Setup} query.

\smallskip \noindent $H_1(\cdot)$ \emph{queries}: $\mathcal {CH}$ keeps a list
$H_1^{list}$ of tuples $(ID_i, m_i, cert_{\mathcal{T}_i}, h_i, coin_i')$. This list
is initially empty. Whenever $\mathcal {A}$ issues a query
$H_1(ID_i,m_i,cert_{\mathcal{T}_i})$, $\mathcal {CH}$ does the following:
\begin{itemize}
  \item If there is a tuple $(ID_i,m_i, cert_{\mathcal{T}_i},h_i,coin_i')$ on
  $H_1^{list}$, return $h_i$ as the answer.
  \item Else, submit $(ID_i,0)$ to $H_0$ and recover the tuple
       $(ID_i,\alpha_{i,0},\alpha_{i,0}',\alpha_{i,1},\alpha_{i,1}',id_{i,0},id_{i,1},coin_i)$ from $H_0^{list}$,
       recover the tuple $(ID_{\mathcal{T}_i},\kappa_i,y_i,cert_{\mathcal{T}_i},$ $coin_{\mathcal{T}_i})$
from $TA^{list}$, flip a coin $coin_i'\in \{0,1\}$ that yields 1 with probability $\delta$ and 0
with probability $1-\delta$. Then do the following:
     \begin{itemize}
       \item If $coin_{\mathcal{T}_i}=coin_i=1$ and $coin_i'=1$, add $(ID_i,m_i,
       cert_{\mathcal{T}_i},h_i,coin_i')$ to $H_1^{list}$ and
       return $h_i=-\alpha_{i,0}'/\alpha_{i,1}'$ as the answer.

       \item Else, randomly select $h_i\in \mathbb{Z}_q^*$, add
       $(ID_i,m_i, cert_{\mathcal{T}_i},h_i,coin_i')$ to $H_1^{list}$ and return
       $h_i$ as the answer.
     \end{itemize}
\end{itemize}

\noindent {\sf Corrupt.LowerLevel} \emph{queries}: On input an identity
$ID_{\mathcal{T}_i}$ of a lower-level TA, $\mathcal {CH}$ first makes
a {\sf LowerLevel.Setup} query on $ID_{\mathcal{T}_i}$, and recovers the
tuple
$(ID_{\mathcal{T}_i},\kappa_i,y_i,cert_{\mathcal{T}_i},coin_{\mathcal{T}_i})$
on $TA^{list}$. If $coin_{\mathcal{T}_i}=0$, $\mathcal {CH}$ returns
$\kappa_i$ as the answer; otherwise, $\mathcal{C}$ aborts.

\smallskip
\noindent {\sf Extract} \emph{queries}: When $\mathcal {A}$ issues
an {\sf Extract} query on $(ID_i,cert_{\mathcal{T}_i})$, the same
answer will be given if the request has been asked before.
Otherwise, $\mathcal {CH}$ recovers
$(ID_{\mathcal{T}_i},\kappa_i,y_i,cert_{\mathcal{T}_i},coin_{\mathcal{T}_i})$
from $TA^{list}$; $\mathcal{C}$ checks whether
$(ID_i,\alpha_{i,0},\alpha_{i,0}',\alpha_{i,1},\alpha_{i,1}',id_{i,0},id_{i,1},coin_i)$
is on $H_0^{list}$; if it is not, $\mathcal {CH}$ submits
$(ID_i,j)$ to $H_0(\cdot)$ to generate such a tuple, where $j=0$ or
1. Finally,
if $coin_i=coin_{\mathcal{T}_i}=1$, $\mathcal {CH}$ aborts;
else if $coin_{\mathcal{T}_i}=0$, it returns
  $(id_{i,0}^{\kappa_i},id_{i,1}^{\kappa_i})$; else it returns $(\psi(g_2^{b\kappa_i\alpha_{i,0}}),\psi(g_2^{b\kappa_i\alpha_{i,1}}))$.

\smallskip
\noindent {\sf Sign} \emph{queries}: On receiving a {\sf Sign} query
on $(ID_i,m_i,cert_{\mathcal{T}_i})$, $\mathcal {CH}$ first queries
$H_0(ID_i,j)$ $(j=0$ or 1), {\sf
LowerLevel.Setup}$(ID_{\mathcal{T}_i})$ and
$H_1(ID_i,m_i,cert_{\mathcal{T}_i})$ if they were not queried
before, then recovers
$(ID_i,\alpha_{i,0},\alpha_{i,0}',\alpha_{i,1},\alpha_{i,1}',id_{i,0},id_{i,1},coin_i)$
from $H_0^{list}$,
$(ID_{\mathcal{T}_i},\kappa_i,y_i,cert_{\mathcal{T}_i},coin_{\mathcal{T}_i})$
from $TA^{list}$ and $(ID_i,m_i,coin_{\mathcal{T}_i},h_i,coin_i')$
from $H_1^{list}$. Finally $\mathcal {CH}$ generates the signature as
follows:
\begin{itemize}
    \item  If $coin_i=coin_{\mathcal{T}_i}=coin_i'=1$, compute and output
    $\sigma_i=\psi(g_2^{b\kappa_i({\alpha_{i,0}-\alpha_{i,1}\alpha_{i,0}'/\alpha_{i,1}'})}).$

    \item Else if $coin_i=coin_{\mathcal{T}_i}=1,coin_i'=0$, abort.

    \item Else, use the {\sf Sign} algorithm to generate the signature,
    since the corresponding private key is known to $\mathcal {CH}$.
\end{itemize}

 Note that, as defined in our security
assumptions, an adversary can only get one signature corresponding to the target identity and lower-level TA.
 Hence, $\mathcal {CH}$ aborts if
 $coin_i=coin_{\mathcal{T}_i}=1,coin_i'=0$.

\smallskip
\noindent \emph{Forgery}: Eventually, $\mathcal {A}$ outputs $l'$
sets of identities $\mathbb{I}_1^*=\{ID_1^*,...,ID_{t_1}^*\}$,
$\mathbb{I}_2^*=\{ID_{t_1+1}^*,...,ID_{t_2}^*\}
,...,\mathbb{I}_{l'}^*=\{ID_{t_{l'-1}+1}^*,...,ID_n^*\}$ enrolled by
$l'$ lower-level TAs with certificates from the set
$\{cert_{\mathcal{T}_1}^*,...,cert_{\mathcal{T}_{l'}}^*\}$, a set of
$n$ messages $\{m_1^*,...,m_n^*\}$ and an aggregate signature
$\Omega^*$. Once $\mathcal {A}$ finishes queries and returns its
forgery, $\mathcal {CH}$ proceeds with the following steps.

For all $i\in\{1,...,n\},j\in \{1,...,l'\}$, $\mathcal {CH}$ finds
$(ID_i^*$, $\alpha_{i,0}^*$, $\alpha_{i,0}'^*$, $\alpha_{i,1}^*$,
$\alpha_{i,1}'^*,$ $id_{i,0}^*$, $id_{i,1}^*$, $coin_i^*)$ on
$H_0^{list}$ and
$(ID_{\mathcal{T}_j}^*,\kappa_j^*,y_j^*,cert_{\mathcal{T}_j}^*,coin_{\mathcal{T}_j}^*)$
on $TA^{list}$. For all $ID_i^*\in \mathbb{I}_j^*$, $\mathcal {CH}$
also recovers the tuples $(ID_i^*, m_i^*,
cert_{\mathcal{T}_j}^*,h_i^*, coin_i'^*)$ from $H_1^{list}$, where
$ID_i^*$ is enrolled by $\mathcal{T}_j$. It is required that there
exists $ID_i^*\in \mathbb{I}_j^*$ such that
$coin_i^*=coin_{\mathcal{T}_j}^*=1$. Without loss of generality, we
assume $i=j=1$. Besides, it is required that for $2\leq i\leq n,
coin_i^*=0$. In addition, the forged aggregate signature must
satisfy
$\hat{e}(\Omega^*,g_2)=\prod_{j=1}^{l'} \hat{e}(\prod_{i\in \mathbb{I}_j'}id_{i,0}^*
      {id_{i,1}^*}^{h_i^*},y_j^*),$ where
$id_{i,0}^*=H_0(ID_i^*,0),id_{i,1}^*=H_0(ID_i^*,1),h_i^*=H_1(ID_i^*,m_i^*,coin_{\mathcal{T}_j}^*)$,
$\mathbb{I}_1'=\{1,...,t_1\}$, $\mathbb{I}_2'=\{t_1+1,...,t_2\}
,...,\mathbb{I}_l'=\{t_{l'-1}+1,...,n\}$. Otherwise, $\mathcal {CH}$
aborts.

Since the forged aggregate signature must satisfy
$\hat{e}(\Omega^*,g_2)=\prod_{j=1}^{l'} \hat{e}(\prod_{i\in \mathbb{I}_j'}id_{i,0}^*
      {id_{i,1}^*}^{h_i^*},y_j^*),$
%      we have
%\begin{eqnarray*}
%&&\hat{e}(\Omega^*,g_2) (\hat{e}(\prod_{i\in \mathbb{I}_1',i\neq 1}id_{i,0}^*
%      {id_{i,1}^*}^{h_i^*},y_1^*)\times\\
%      && \prod_{j=2}^{l'} \hat{e}(\prod_{i\in \mathbb{I}_j'}id_{i,0}^*
%      {id_{i,1}^*}^{h_i^*},y_j^*))^{-1}\\
%&&=\hat{e}(id_{1,0}^* {id_{1,1}^*}^{h_1^*},y_1^*).
%\end{eqnarray*}
and
$id_{1,0}=g_1^{\alpha_{1,0}}{g_1^a}^{\alpha_{1,0}'},id_{1,1}=g_1^{\alpha_{1,1}}{g_1^a}^{\alpha_{1,1}'}$,
for all $i\in \{2,...,n\}$,
$id_{i,0}^*=g_1^{\alpha_{i,0}^*},id_{i,1}^*=g_1^{\alpha_{i,1}^*}$,
we have
\begin{eqnarray*}
g_1^{ab}=(\Omega^*
(\prod_{j=2}^{l'}\prod_{i\in \mathbb{I}_j'}\psi(y_j^*)^{-\sum_{i\in \mathbb{I}_j'}(\alpha_{i,0}^*+h_i^*\alpha_{i,1}^*)})\times\\
      \psi({y_1^*}^{-\sum_{i=1}^{t_1}(\alpha_{i,0}^*+h_i^*\alpha_{i,1}^*)}))^{\frac{1}{\kappa_1^*(\alpha_{1,0}'^*+h_1^*\alpha_{1,1}'^*)}}.
\end{eqnarray*}

To complete the proof, we shall show that $\mathcal {CH}$ solves the
given instance of the co-CDH problem with probability at least
$\epsilon'$. First, we analyze the three events needed for $\mathcal
{C}$ to succeed:

\begin{itemize}
  \item $\Sigma$1: $\mathcal {CH}$ does not abort as a result of any of
  $\mathcal {A}$'s {\sf Corrupt.LowerLevel}, {\sf Extract} and {\sf Sign}
queries.
  \item $\Sigma$2: $\mathcal {A}$ generates a valid and nontrivial aggregate signature
forgery.
  \item $\Sigma$3: $\Sigma$2 occurs, $coin_1^*=coin_{\mathcal{T}_1}=1,coin_1'^*=0$ and for $2\leq i\leq n, coin_i^*=0$.
\end{itemize}

$\mathcal {CH}$ succeeds if all of these events happen. The
probability $\Pr[\Sigma1\wedge \Sigma2\wedge \Sigma3]$ can be
decomposed as $\Pr[\Sigma1\wedge \Sigma2\wedge \Sigma3]
= \Pr[\Sigma1]\Pr[\Sigma2|\Sigma1]\Pr[\Sigma3|\Sigma1
\wedge \Sigma2]$.

\medskip\noindent
\textbf{Claim 1.} The probability that $\mathcal {CH}$ does not abort
as a result of $\mathcal {A}$'s {\sf Corrupt.LowerLevel}, {\sf
Extract} and {\sf Sign} queries is at least
$(1-\delta)^{q_C+q_E+q_S}$. Hence we have $\Pr[\Sigma1] \geq
(1-\delta)^{q_C+q_E+q_S}$.

\emph{Proof:}
For a {\sf Corrupt.LowerLevel} query, $\mathcal {CH}$ will abort iff
$coin_{\mathcal{T}_i}=1$. It is easy to see that the probability
that $\mathcal {CH}$ does not abort is $1-\delta$. Since $\mathcal
{A}$ can make at most $q_C$ times {\sf Corrupt.LowerLevel} queries,
the probability that $\mathcal {CH}$ does not abort as a result of
$\mathcal {A}$'s {\sf Corrupt.LowerLevel} queries is at least
$(1-\delta)^{q_C}$.

For an {\sf Extract} query, $\mathcal {CH}$ will abort iff
$coin_i=coin_{\mathcal{T}_i}=1$. It is easy to see that the
probability that $\mathcal {CH}$ does not abort for an {\sf Extract}
query is $1-\delta^2> 1-\delta$. Since $\mathcal {A}$ can make at
most $q_E$ times {\sf Extract} queries, the probability that
$\mathcal {CH}$ does not abort as a result of $\mathcal {A}$'s {\sf
Extract} queries is at least $(1-\delta)^{q_E}$.

When $\mathcal {CH}$ receives a {\sf Sign} query, he will abort iff
$coin_i=coin_{\mathcal{T}_i}=1,coin_i'=0$ happen. So for a {\sf Sign} query, the
probability that $\mathcal {CH}$ does not abort is
$1-\delta^2(1-\delta)>1-\delta$. Since $\mathcal {A}$ makes at most
$q_S$ times {\sf Sign} queries, the probability that
 $\mathcal {CH}$ does not abort as a result of $\mathcal
{A}$'s {\sf Sign} queries is at least $(1-\delta)^{q_S}$.

Overall, we have $\Pr[\Sigma1] > (1-\delta)^{q_C+q_E+q_S}$.

\medskip
\noindent \textbf{Claim 2.} $\Pr[\Sigma2|\Sigma1]\geq \epsilon$.

\emph{Proof:}
If $\mathcal {CH}$ does not abort, then $\mathcal {A}$'s
view is identical to its view in the real attack. Hence,
$\Pr[\Sigma2|\Sigma1]\geq \epsilon$.

\medskip\noindent
\textbf{Claim 3.} The probability that $\mathcal {CH}$ does not abort
after $\mathcal {A}$ outputting a valid and nontrivial forgery is at
least $\delta(1-\delta)^n$. Hence $\Pr[\Sigma3|\Sigma1\wedge
\Sigma2]\geq \delta(1-\delta)^n$.

\emph{Proof:}
Events $\Sigma1$ and $\Sigma2$ have occurred, and $\mathcal {A}$
has generated a valid and nontrivial forgery
$(ID_1^*,...,ID_n^*;m_1^*,...,m_n^*,\Omega^*)$. $\mathcal {CH}$ will
abort unless $\mathcal {A}$ generates a forgery such that there
exists an $i\in \{1,...,n\}$ such that
$coin_1^*=coin_{\mathcal{T}_1}^*=1,coin_1'^*=0$, and for  $2\leq
i\leq n$, $coin_i^*=0$. Therefore, $\Pr[\Sigma3|\Sigma1\wedge
\Sigma2]\geq \delta^2(1-\delta)^n$.

\medskip
%JOSEP3. Totally -> In total
\indent In total, we have
$\epsilon'=\Pr[\Sigma1\wedge \Sigma2\wedge \Sigma3]
>(1-\delta)^{q_C+q_E+q_S}\delta^2(1-\delta)^n \epsilon
\geq \frac{4}{e^2(q_C+q_E+q_S+n+2)^2}\epsilon,$
where $e$ is Euler's constant.

\section{Conclusion}\label{Conclusion}

We have proven that our MTA-OTIBAS concrete scheme is existentially
unforgeable against adaptively chosen-message attacks in
the random oracle model under the co-CDH problem
assumption.
%\section*{Acknowledgments}
%This work was supported by the NSFC under Grants 61202465, 61021004, 61103222,
%1173154, 61003214, 61272501 and 61370190; the Shanghai NSF under Grant 12ZR1443500; the Shanghai Chen Guang Program
%(12CG24); the Science and Technology Commission of Shanghai Municipality under grant 13JC1403500; the National Key Basic Research Program (973 program) through Project
%2012CB315905;
%the Beijing NSF through Project 4132056; EU FP7 under Projects ``DwB'' and ``Inter-Trust'';
%the Government of Catalonia under grant 2014 SGR 537; the Research Fund for the Doctoral Program of Higher Education
%of China under Grant No. 20110076120016; the Fundamental Research Funds for the Central Universities of China; the Research Funds of Renmin University
%of China (No.14XNLF02). J. Domingo-Ferrer was supported in
%part as an ICREA-Acad\`emia researcher by the Government of Catalonia.

%%%%%%%%%%%%%%%%%%%%%%%%%%%%%%%%%%%%%%%%%%%%%%%%%%%%%%%%%%%%%%%%%%%%%%%%%

\end{document}